# Versatile optimization-based speed-up method for autofocusing in digital holographic microscopy

JULIANNA WINNIK,[1,5] DAMIAN SUSKI,[2] PIOTR ZDAŃKOWSKI,[1] LUIZA STANASZEK,[3] VICENTE MICÓ,[4] AND MACIEJ TRUSIAK[1,6]

[1]*Warsaw University of Technology, Institute of Micromechanics and Photonics, 8 Sw. A. Boboli St., 02-525 Warsaw, Poland*
[2]*Warsaw University of Technology, Institute of Automatic Control and Robotics, 8 Sw. A. Boboli St., 02-525 Warsaw, Poland*
[3]*NeuroRepair Department, Mossakowski Medical Research Centre, Polish Academy of Sciences, 5 Pawińskiego St., 02-106, Warsaw, Poland*
[4]*Departamento de Óptica y de Optometría y Ciencias de la Visión. Facultad de Física. Universidad de Valencia. C/ Doctor Moliner 50, 46100 Burjassot, Spain*
[5]*julianna.winnik@pw.edu.pl*
[6]*maciej.trusiak@pw.edu.pl*

**Abstract:** We propose a speed-up method for the in-focus plane detection in digital holographic microscopy that can be applied to a broad class of autofocusing algorithms that involve repetitive propagation of an object wave to various axial locations to decide the in-focus position. The classical autofocusing algorithms apply a uniform search strategy, i.e., they probe multiple, uniformly distributed axial locations, which leads to heavy computational overhead. Our method substantially reduces the computational load, without sacrificing the accuracy, by skillfully selecting the next location to investigate, which results in a decreased total number of probed propagation distances. This is achieved by applying the golden selection search with parabolic interpolation, which is the gold standard for tackling single-variable optimization problems. The proposed approach is successfully applied to three diverse autofocusing cases, providing up to 136-fold speed-up.

## 1. Introduction

Digital holographic microscopy (DHM) [1-3] enables registration of an optical field that has been disturbed, i.e., refracted or reflected, by a microscale sample and thus enables gaining valuable information about its features such as the geometry, refractive index and absorptive properties. The true power of DHM comes from numerical refocusing tools [4-7] that enable algorithmic simulation of propagation of the light wave in space. Most importantly, numerical refocusing enables enhancing the image sharpness by propagating the captured optical field to the in-focus location. However, to fulfill this task one needs to know the required propagation distance, which in the general case is not known a priori or is known with the limited, unsatisfactory accuracy.

The problem can be addressed with powerful, holographic autofocusing [8-27]. It is worth noticing that the application of autofocusing is not limited to the standard focus enhancement in the conventional DHM. It can be applied to a wide range of tasks, e.g., particle localization [28] also in lensless DHM [29-32], time-lapse study of dynamic objects [33], tilted image plane detection [34], digital holography of macroscale samples [35], rotation errors correction in holographic tomography [36-40] and accurate shape recovery [41].

The in-focus plane detection can be performed using various autofocusing approaches. The most popular class of the autofocusing algorithms utilizes repetitive propagation of the object wave to multiple positions along the optical axis [8-27]. In each location, the defocus of the object wave is quantified using one of the available focus metrics. The plane with the minimum defocus value is assumed to be the in-focus one. Recently, also another class of autofocusing algorithms has emerged that utilizes deep learning [42-45]. The drawback of these methods is a requirement for a large learning set. Here we focus on the former, most popular autofocusing approach that applies repetitive propagation. The major disadvantage of this solution is the high computational cost. The issue has been addressed with the autofocusing acceleration via hologram downsampling [46,47], efficient implementation of the autofocusing procedure on graphics processors [48] and a two-step approach with preliminary and precise autofocusing stages [8,49].

In this paper we claim that the high computational cost of autofocusing comes mostly from the uniform search strategy, i.e., the autofocusing algorithms investigate numerous, equidistant planes, which is insufficient and leads to heavy computational overhead. We address this issue by proposing a versatile speed-up method for autofocusing in DHM that can work with both any focus metrics and DHM configuration. First, we formulate the autofocusing task as the optimization problem. Then, we replace the uniform search strategy with a suitable algorithmic tool for single-variable optimization problems, that is the golden selection search with parabolic interpolation (GSS-PI) [50]. GSS-PI, at each iteration, skillfully decides the next axial position to investigate, which results in a reduced total number of probed axial locations. Importantly, the acceleration does not affect the autofocusing accuracy.

The proposed speed-up approach is successfully verified in application to three diverse autofocusing cases: 1) focus enhancement of an alive mesenchymal stem cell in a grating-assisted, common-path DHM system; 2) detection of the microbead location using lensless in-line DHM and dark focus metric; 3) in-focus plane detection of USAF resolution test target using an autofocusing algorithm based on two-directional illumination and DHM in Mach-Zehnder configuration. In all investigated cases, the proposed approach enabled substantial speed-up of the autofocusing procedure.

## 2. Conventional autofocusing algorithms

DHM allows acquiring complete information about the scalar object wave, i.e., its amplitude and phase. However, in many holographic systems, the hologram acquisition plane does not coincide with the image plane, which results in a blurry, often useless reconstruction. The defocused registration conditions may be inherent to the working principle of a given holographic setup, e.g., [51-53]. Nonetheless, even in the image plane holographic configuration, the defocus may arise from the sample dynamic character, the setup misalignments or the need to image to its best in-focus plane a thick sample (thicker than the depth of field of the imaging system). In either of the cases, the defocusing problem can be addressed with numerical propagation that enables computational refocusing of the object wave to the in-focus plane. This, however, requires precise knowledge about the distance between the hologram acquisition plane and the in-focus plane.

The described problem is schematically presented in Fig. 1. In order to find the in-focus distance $z_f$, the object wave $u_{z=0}$, registered at plane $z = 0$, is repetitively propagated on various distances $z$, which can be done using, e.g., angular spectrum method [4]:

$$\tilde{u}_{z=0}(f_x, f_y) = \iint u_{z=0}(x, y) \exp\left[-i2\pi(f_x x + f_y y)\right] dx dy, \tag{1}$$

$$\tilde{u}_z(f_x, f_y) = \tilde{u}_{z=0}(f_x, f_y) \exp\left[i2\pi z \sqrt{\left(\frac{n}{\lambda}\right)^2 - f_x^2 - f_y^2}\right], \tag{2}$$

$$u_z(x, y) = \iint \tilde{u}_z(f_x, f_y) \exp\left[i2\pi(f_x x + f_y y)\right] df_x df_y. \tag{3}$$

In Eqs (1)-(3) tilde denotes Fourier transform, $i$ is an imaginary unit, $(f_x, f_y)$ are spatial frequencies, $\lambda$ stands for the light wavelength, $n$ is refractive index of the medium in which the propagation takes place.

Crucially, in the autofocusing algorithms the investigated locations are uniformly distributed with a step $\Delta z$. At each plane, a focus metrics of choice is applied to evaluate the focusing conditions. The plane with the minimum defocus value is assumed to be the in-focus one.

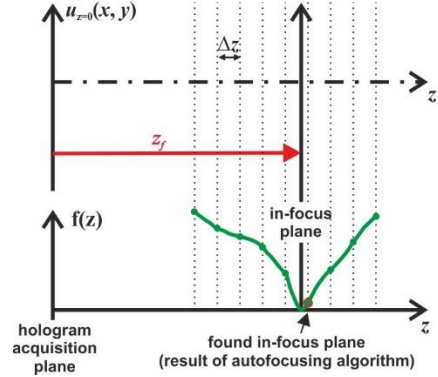

Fig. 1. Working principle of the conventional autofocusing algorithm;
$u$ – optical field, $f$ – defocus function, $z$ – propagation distance; $\Delta z$ –
search step, $z_f$ – optimal propagation distance.

## 3. Autofocusing utilizing golden selection search with parabolic interpolation

### 3.1 Formulation of the optimization task

From the computational point of view, we aim at solving the following optimization problem:

$$\min_{z \in \mathbb{R}} f(z)$$
$$s.t.\ z \in [z_{\min}, z_{\max}] \tag{4}$$

where $f(z)$ is the defocus metric calculated for each $z$ based on the numerically refocused object wave $u_z$ and $[z_{\min}, z_{\max}]$ is an arbitrarily chosen interval in which the in-focus distance $z_f$ is searched. We assume that $f(z)$ is a continuous function of $z$. In our convention $f(z)$ is a degree of defocus therefore if the given focus metric is maximized at the in-focus plane, e.g. [10, 29], then we take negative of this metric.

It should be noted that numerical algorithms for solving the considered optimization problem can only guarantee to find the global optimum if the function $f(z)$ is unimodal, i.e., it possesses only one local minimum on the search interval $[z_{\min}, z_{\max}]$ [50]. If the function $f(z)$ is multimodal the numerical algorithms may converge to the local optimum that is not actually the global minimum. This is a general problem in numerical optimization. To increase the chance of finding the global optimum, some heuristics can be applied, e.g., splitting the original interval of search into several subintervals and searching the minimum in each of them separately.

### 3.2 Golden selection search with parabolic interpolation

The state-of-the-art algorithm for solving the considered optimization problem [Eq. (4)] comes from the combination of two algorithms, the golden section search (GSS) and parabolic interpolation (PI), proposed by Brent [50]. The GSS-PI results in an efficient and robust method for solving the considered single-variable optimization problem. For the completeness of the proposed paper, the advantages and drawbacks of GSS and PI algorithms will be now briefly discussed together with a short description of each of them.

The idea of GSS is to shrink the search interval at each iteration based on values of $f(z)$ at four points [50], see Fig. 2. The points $z^i_{\min}$ and $z^i_{\max}$ denote the left and right end of the search interval at $i$-th iteration. At the first iteration we take $z^i_{\min} = z_{\min}$ and $z^i_{\max} = z_{\max}$. Next, we take two points inside the search interval $z_1^i$ and $z_2^i$, which are placed symmetrically with respect to the interval ends:

$$z_1^i = z_{min}^i + (1-\phi) \cdot \left(z_{max}^i - z_{min}^i\right), \tag{5}$$

$$z_2^i = z_{min}^i + \phi \cdot \left(z_{max}^i - z_{min}^i\right), \tag{6}$$

where $\phi = \left(\sqrt{5}-1\right)/2 \approx 0{,}618033$. The inverse of $\phi$ is called the golden ratio, from which the algorithm takes its name.

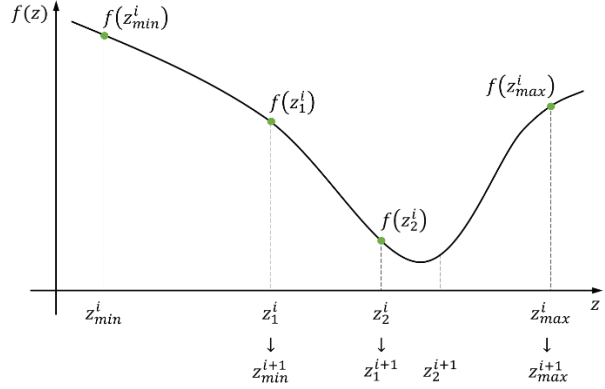

Fig. 2. Illustration of the GSS algorithm: $f$ – defocus function; $z$ – propagation distance; $z^i_{min}$, $z^i_1$, $z^i_2$, $z^i_{max}$ – investigated points at the current iteration ($z^i_{min}$ and $z^i_{max}$ denote ends of the current search interval; $z^i_1$ and $z^i_2$ are selected on the basis of the golden ratio).

For the unimodal function, if the following condition holds:

$$f(z_{min}^i) > f(z_1^i) > f(z_2^i), \tag{7}$$

then we know that on the interval $\left[z_{min}^i, z_1^i\right]$ the function $f(z)$ is decreasing thus we can drop that interval at the next iteration, see Fig. 2. The new search interval will be defined by $z_{min}^{i+1} = z_1^i$ and $z_{max}^{i+1} = z_{max}^i$. The advantage of taking $\phi = \left(\sqrt{5}-1\right)/2$ instead of any other value, is that $z_1^{i+1}$ is exactly equal to $z_2^i$ so we only need to calculate $f(z)$ at one new point $z_2^{i+1}$, which saves the computational effort. Analogically, if

$$f(z_1^i) < f(z_2^i) < f(z_{max}^i) \tag{8}$$

then we know that on the interval $\left[z_2^i, z_{max}^i\right]$ the function $f(z)$ is increasing and we can drop that interval at the next iteration. The new interval will be given by $z_{min}^{i+1} = z_{min}^i$ and $z_{max}^{i+1} = z_2^i$. We also have $z_2^{i+1} = z_1^i$ thus we only need to calculate $f(z)$ at one new point $z_1^{i+1}$.

We repeat the above procedure at subsequent iterations. The length of the search interval decreases by a factor of $\phi$ at each iteration and we stop the procedure when $z^i_{max} - z^i_{min}$ is less than a given declared tolerance *tol*.

If $f(z)$ is unimodal on the original search interval, then GSS guarantees to find the optimum value within the declared tolerance [50]. If $f(z)$ is multimodal, we do not have such a warranty. Instead, we shrink the search interval from iteration to iteration and at some iteration, the actual search interval contains only one local minimum. From that iteration, GSS will search for that local minimum and reach it, but we have no warranty that the local minimum is the global minimum of the original optimization task. From this feature, it follows that the method, in its original form, is not suitable for the autofocusing tasks with multiple focal locations [54].

The idea behind PI is to subsequently build the parabolic interpolations of the function $f(z)$ based on three points $z_1^i$, $z_2^i$ and $z_3^i$, [50], see Fig. 3(a). Next, we find the minimum of the parabolic interpolation function at the point $z_{PI}^i$. We then go to the next iteration taking the new values $z_1^{i+1} = z_2^i$, $z_2^{i+1} = z_3^i$ and $z_3^{i+1} = z_{PI}^i$. The advantage of PI is its fast convergence provided that we are close enough to the minimum of the function $f(z)$ [50]. The parabolic interpolations give a better and better approximation of the $f(z)$ function with each iteration, see Figs 3(b). The drawback of the PI algorithm is that if we are far from the optimal point, the parabolic interpolations may behave poorly [50].

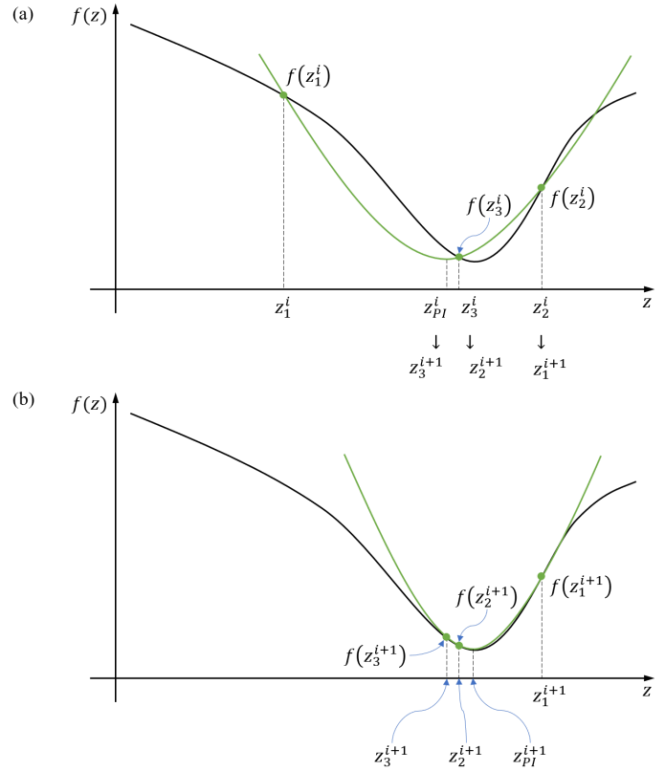

Fig. 3 Illustration of the PI algorithm at (a) $i$-th and (b) ($i$+1)-th iteration (away and close to the minimum of the defocus function $f(z)$, for (a) and (b), respectively); $z^i{}_1$, $z^i{}_2$, $z^i{}_3$ – the propagation distances used for parabolic interpolation at the current iteration; $z^i{}_{PI}$ – minimum of the fitted parabolic function.

The idea behind GSS-PI is to combine the advantages of GSS (guaranteed convergence at least to some local minimum) and PI (fast convergence). Whenever possible PI is used, but when the performance of PI is poor, e.g., $z_{PI}^i$ lies outside the current search interval, the algorithm switches to pure GSS.

The GSS-PI algorithm has been implemented in the function *fminbound* from the open-source Python *scipy* library [55]. That implementation has been utilized in our work.

In comparison to the uniform search method, GSS-PI finds the minimum of the defocus function at a much lower computational cost. However, a drawback of GSS-PI is that the search interval must be chosen with care to avoid omitting the minimum by the algorithm. The reliability of GSS-PI can be improved with the two-step autofocusing procedure proposed in

[8]. In this method, first, the uniform search strategy with a large step $\varDelta z$ is used to obtain a rough profile of the defocus function on the original search interval. This enables defining a new, shrunk search interval, which is investigated further on using a much smaller step $\varDelta z$, to provide high precision. This two-step strategy offers a trade-off between the computational time and the reliability. GSS-PI has a potential to decrease the computational effort of the second step of this strategy, without affecting the reliability.

## 4. Experimental results

The proposed speed-up approach is tested with three, diverse autofocusing cases: (1) focus enhancement of a stem cell in recently proposed, grating-assisted DHM [56], (2) particle localization in lensless, in-line DHM, and (3) focus correction of USAF resolution target in DHM in Mach-Zehnder configuration and two-directional illumination. It is worth noticing that our method can be applied to any complex amplitude image provided by any other DHM architecture.

*4.1 Autofocusing case 1: hologram of a stem cell captured with a grating-assisted DHM*

First autofocusing case is concerned with the recently proposed, common-path DHM system [56], where the object wave, after imaging with a microscope optical setup, is incident on a diffraction grating. The $+1^{st}$ and $-1^{st}$ diffraction orders interfere producing an off-axis hologram. The system operates in the total-shear regime. The samples for our experiment are alive mesenchymal stem cells. The hologram was registered using a laser diode with a light wavelength $\lambda = 635$ nm and the image sensor with the pixel pitch of 1.85 μm. The parameters of the optical system are magnification $M = -20$ and the numerical aperture $NA = 0.75$. The hologram was reconstructed using the Fourier transform method [57]. The reconstructed object wave amplitude is presented in Fig. 4, where the red line indicates the region of interest containing a single cell, which was used for the evaluation of the focusing conditions.

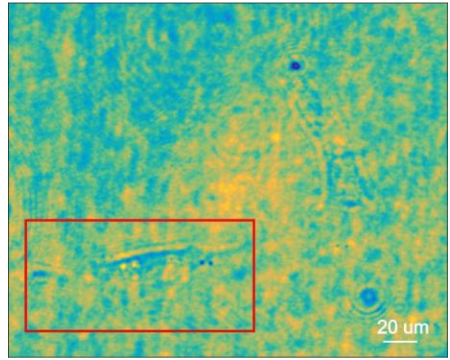

Fig. 4. Full field of view of the object wave amplitude with the indicated region of interest used for evaluation of the focusing conditions; the sample is a mesenchymal stem cell; the object wave was captured with a grating-assisted DHM system [56].

The considered system, typically, works in the image plane holography regime, which means that the hologram acquisition plane is optically conjugated with a sample, thus no defocus occurs. However, in practice, a small defocus may arise from, e.g., instabilities of the setup, or as in this case, movement of a living specimen. Therefore, for this test, the axial range of search was set symmetrically around the hologram acquisition plane, $z \in [-100$ μm, $100$ μm$]$.

Generally, the choice of the axial scanning step is dictated by parameters of the system such as $NA$, wavelength, magnification (all influencing the depth of field) as well as the character of the sample and the required accuracy. In our case, the search step was set to $\varDelta z = 1$ μm. This value was also taken as tolerance for GSS-PI, which ensured equal accuracy of both autofocusing approaches and thus facilitated their comparison. The chosen value of *tol* = *Δz* is slightly smaller than the depth of field (DoF) of the considered DHM system (here DoF ≈ 1.5 µm [58]), which, in theory, ensures that the autofocusing will bring the data into focus.

The stem cell is treated here as a pure phase object, which is expected to be invisible when focused. Therefore, the defocus value can be evaluated with a variance of the amplitude of the object wave [34,36,38] that was numerically refocused on distance *z*:

$$f(z) = \mathrm{var}(|u_z|). \tag{9}$$

The numerical refocusing is handled with the angular spectrum method using the whole field of view; however, the focus measure is evaluated only in the region of interest surrounding the sample, (Fig. 4), which enhances the autofocusing accuracy. The variance *var* is given by:

$$\mathrm{var}(u) = \frac{1}{P}\sum_{p=1}^{P}(u_p - \bar{u})^2, \tag{10}$$

where *p* denotes the pixel index and $\bar{u}$ is the average signal value.

The results of the autofocusing, performed with both uniform search strategy and GSS-PI, are shown in Fig. 5(a). Both autofocusing algorithms pointed to almost the same axial location (Tab. 1). The conventional approach investigated 200 axial locations, while GSS-PI looked up only 10 planes, providing substantial 20-fold speed-up.

The zoomed amplitude and phase of the object wave in the hologram registration plane and in the determined in-focus plane (according to GSS-PI) are presented in Figs 5(b)-5(c) and Figs 5(d)-5(e), respectively. Almost uniform amplitude distribution in Fig. 5(d) indicates the successful defocus correction. It can be observed that the autofocusing improved visibility of the cell structure in the phase image, Fig. 5(e).

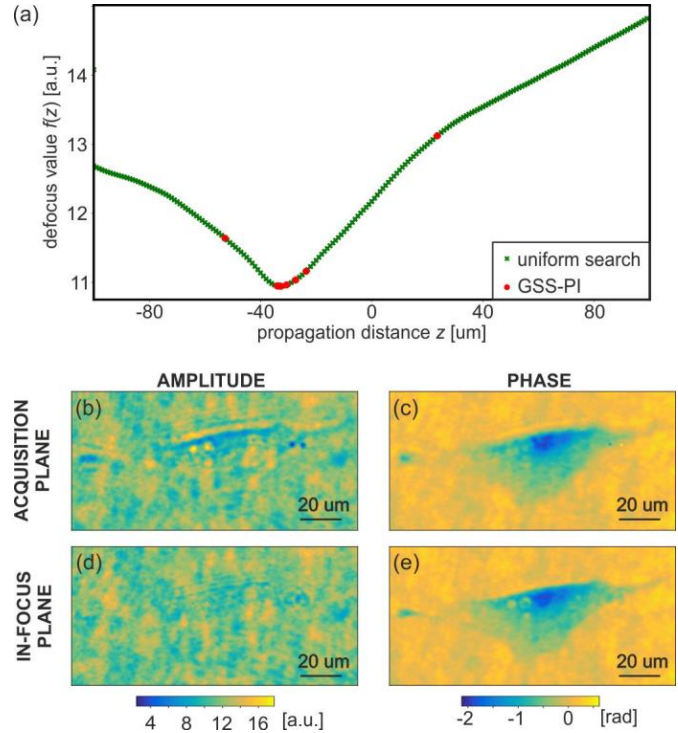

Fig. 5. Autofocusing results for the mesenchymal stem cell: (a) defocus function evaluated with a variance of amplitude of the object wave; amplitude (b, d) and phase (c, e) of the object wave in the hologram acquisition plane (b, c) and in the found in-focus plane (d, e) (zoomed area).

Table 1. Results of autofocusing for the mesenchymal stem cell.

|  | Uniform search | GSS-PI |
|---|---|---|
| Number of investigated planes | 200 | 10 |
| Found in-focus location [μm] | -33.0 | -33.1 |

*4.2 Autofocusing case 2: hologram of microbeads captured with lensless in-line DHM*

The second analyzed autofocusing case concerns the particle localization using a lensless in-line DHM in Gabor configuration [29]. In our experiment, the samples are polystyrene beads with a diameter of 90 μm immersed in water and inserted in a counting chamber of 100 μm thickness (thus, there is essentially a single plane containing all the beads). The hologram was registered using an image sensor (2048 X 2048 pixels, pixel pitch of 5.5 μm) that was placed in a distance of 300 mm from the light source, here a fiber-coupled laser diode with $\lambda = 450$ nm. The registered hologram is presented in Fig. 6. The red rectangle indicates the region of interest with a single selected bead that was used for the autofocusing procedure.

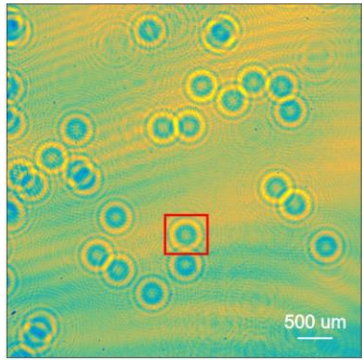

Fig. 6. Full field of view of the object wave amplitude with the indicated region of interest (red line) that was used for evaluation of the focusing conditions; the sample is a microbead with a diameter of 90 μm; the object wave was captured with lensless in-line DHM.

The working principle of lensless DHM imposes defocused registration conditions, i.e., the sample is placed at some distance from the image sensor, which when combined with diverging beam illumination, enables achieving optical magnification. Therefore, in this autofocusing case, the axial range of search was set to a large area in the front of the image sensor: $z \in$ [-300 mm, 0]. We set $\Delta z = tol = 0.2$ mm. The evaluation of the focus condition was performed with the dark focus metric [29], which is a robust indicator of the overall sharpness for objects with mixed amplitude-phase properties. Dark focus calculates the gradient variance of the numerically generated dark field $u_z^d$:

$$f(z) = -\sqrt{\text{var}\left(\Delta \left|u_z^d\right|\right)}. \tag{11}$$

Note that in our convention the in-focus plane is always indicated with a minimum of *f(z)* thus Eq. (11) expresses the negative of the dark focus metric.

The results of the autofocusing, performed with the uniform search strategy and GSS-PI, are presented in Fig. 7(a). In the analyzed case GSS-PI provided tremendous, 136-fold autofocusing speed-up, without scarifying the accuracy (Tab. 2). This fact provides evidence that the broader the search range, the larger the computational speed-up ensured by GSS-PI in comparison with the uniform search strategy. Thus, results presented in Fig. 7(a) promote GSS-PI for high-throughput large volume 3D holographic imaging.

The comparison of the object wave amplitudes and phases before and after autofocusing is shown in Fig. 7(b)-7(e). The diffraction fringes in the amplitude image, Fig. 7(b), indicate a large defocus of the initial data. From Fig. 7(c) it can be noticed that the initial phase has been assumed to be uniform, which complies with the all-intensity hologram reconstruction method for lensless DHM [29]. It is worth noting that the applied reconstruction approach does not deal with the twin image problem. After the application of GSS-PI, the optical field was refocused to the found in-focus location at -100.74 mm, which resulted in the sharp reconstruction, Figs 7(d) and 7(e). It can be observed that the transparent bead is imaged in amplitude as a dark, sharp circle with a bright central spot, Fig. 7(d). This is related to small NA of the given in-line lensless DHM system, which filters out the rays that are strongly refracted on the high slope areas of the investigated microsphere.

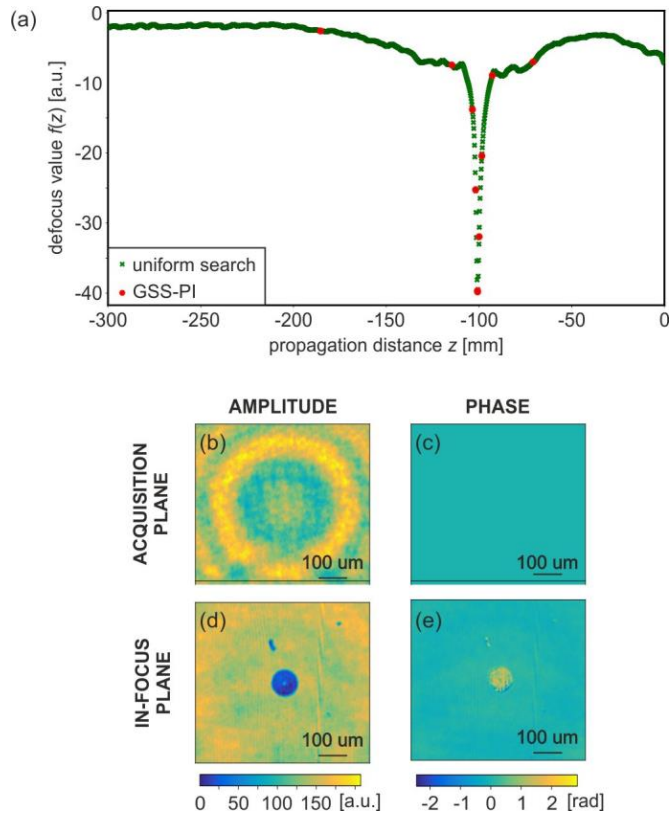

Fig. 7. Autofocusing results for the microbead: (a) defocus function evaluated with the dark focus metric; Amplitude (b, d) and phase (c, e) of the object wave in the hologram acquisition plane (b, c) and in the detected in-focus plane (d, e) (zoomed area).

**Table 2. Results of autofocusing for the microbead.**

|  | *Uniform search* | *GSS-PI* |
|---|---|---|
| *Number of investigated planes* | 1500 | 11 |
| *Found in-focus location [mm]* | -100.80 | -100.74 |

### 4.3 Autofocusing case 3: two holographic views of the USAF resolution test target captured with DHM in Mach-Zehnder configuration

Two previously discussed autofocusing methods were concerned with the focus metrics that quantify the defocus basing on a single object wave. In the third, last example we investigate a different autofocusing approach that looks for the in-focus plane by analyzing the interdependence of two object waves that correspond to various illumination directions [15]. The off-axis propagation directions of the waves invoke transverse, mutual displacement of the sample images in the defocus planes. Thus, the in-focus location can be found by looking for minimum variance between the waves' amplitudes:

$$f(z) = \text{var}\left(|u_z^1| - |u_z^2|\right). \tag{12}$$

The discussed autofocusing method was applied to holographic tomography [40], where it addressed the key problem of rotation errors and related data defocusing. Here, we demonstrate the possibility of acceleration of this autofocusing method using GSS-PI. In our experiment, two holograms of the USAF resolution test target were taken with off-axis DHM in Mach-Zehnder configuration and two-directional tilted beam illumination (the illumination vectors lie in the horizontal plane and form +/-13.5° angle with the optical axis) [40]. The system applied a He-Ne laser with $\lambda$ = 632.8 nm, a microscope optical system with $M$ = -19.5 and $NA$ = 0.42, and an image sensor of size 2456 ✕ 2058 and a pixel pitch of 3.45 μm. The holograms were reconstructed using Fourier transform method [57]. The amplitudes of the reconstructed object waves are shown in Fig. 8.

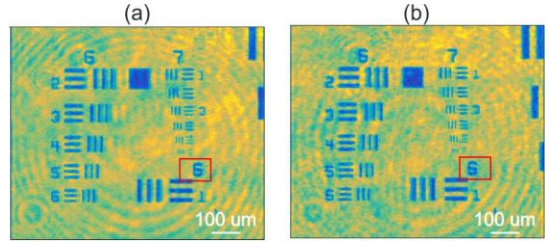

Fig. 8. Full field of view of two object wave amplitudes corresponding to illumination at (a) +13.5° and (b) -13.5°; red rectangles indicate a region of interest that was used for evaluation of the focusing conditions; the sample is USAF resolution test target; the object waves were captured with DHM in Mach-Zehnder off-axis configuration.

The considered DHM system operates in the image plane holography regime, where a small defocus may occur due to the setup instability or inaccurate placing of the sample. Therefore, the axial search was defined as a small region around the hologram acquisition plane, i.e., $z \in$ [-50 μm, 50 μm]. We set $\Delta z = tol = 2$ μm (here the search step is larger than in Sec. 4.1 due to smaller NA and thus larger DoF of the considered DHM system, here DoF ≈ 2.5 μm [58]). The obtained autofocus results are presented in Fig. 9.

In this case, GSS-PI provided a 7-fold acceleration of the autofocusing without decreasing its accuracy (Tab. 3). The smaller gain in computational complexity comes from a relatively sparse sampling of the autofocus curve due to large DoF of the employed optical system.

Figures 9(b)-9(e) compares the amplitudes of the pair of the object waves in the hologram registration plane, Figs 9(b) and (c), and in the in-focus plane, Figs 9(d) and (e). In this case, we did not include the phase images for the sake of a concise presentation. One can notice the transverse, mutual shift of the images in the original plane, Figs 9(b) and (c), which indicates the defocused registration conditions. The shift is removed after propagation to the in-focus plane at $z = 21.70$ μm.

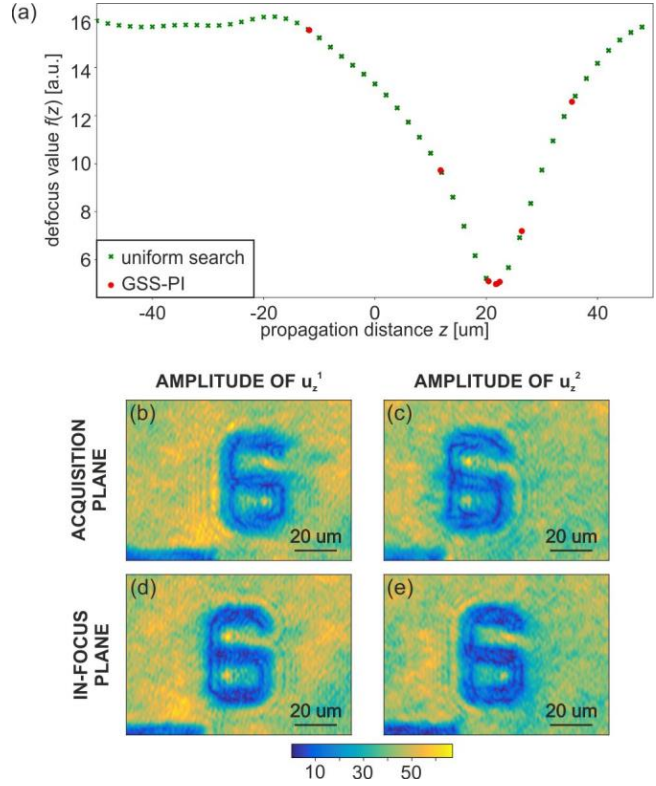

Fig. 9. Autofocusing results for USAF resolution test target: (a) defocus function evaluated with the autofocusing method proposed in [15]; zoomed areas of amplitudes of two object waves, corresponding to different illumination directions: (b, d) +13.5°, (c, e) -13.5°, in the hologram acquisition plane (b, c) and in the found in-focus plane (d, e).

**Table 3. Results of autofocusing for the USAF resolution test target.**

|  | Uniform search | GSS-PI |
|---|---|---|
| Number of investigated planes | 50 | 7 |
| Found in-focus location [μm] | 22.00 | 21.70 |

### 4.4 Summary of the achieved speed-up

The achieved autofocusing speed-up is summarized in Tab. 4, in which we express the computational gain of the proposed algorithm as a ratio of numbers of probed locations in the uniform search strategy and GSS-PI. Using this criterion, the largest, 136.4-fold computation gain was obtained for the autofocusing case 2, i.e., particle localization in lensless DHM. This tremendous computational gain was caused by a very large axial search range, which resulted in numerous probed axial locations for the uniform search strategy. In the case of the image plane holographic configurations, the achieved speed-up was 20-fold and 7.1-fold for autofocusing cases 1 and 3, respectively. The smaller acceleration gain for case 3 comes from the relatively large axial search step $\Delta z$. Generally, the autofocusing acceleration with GSS-PI is expected to be larger for more complex autofocusing problems (wider search range and/or higher required autofocusing accuracy).

Table 4 also includes computational times of the autofocusing procedures for all analyzed cases. Both autofocusing algorithms were implemented in Python and computed on Intel Core

i7-7700HQ 2.80 GHz equipped with 32 GB of RAM. The obtain accelerations comply with the theoretical gain related to the reduced number of probed locations.

Lastly, in all analyzed cases, the difference between the found in-focus locations for the uniform search strategy and GSS-PI was within the declared tolerance *tol*.

**Table 4. Summary of the computational gain of GSS-PI**

| Autofocus case | Number of f(z) evaluations | | Gain | Computational time [s] | | Gain |
|---|---|---|---|---|---|---|
| | uniform search | GSS-PI | | uniform search | GSS-PI | |
| 1 | 200 | 10 | 20 | 236.7 | 12.4 | 19.1 |
| 2 | 1500 | 11 | 136.4 | 1144.3 | 8.6 | 133.1 |
| 3 | 50 | 7 | 7.1 | 156.8 | 22.5 | 7.0 |

## 5. Conclusions

In this paper, we proposed a versatile acceleration method for autofocusing in DHM that can be applied to various focus metrics and DHM configurations. The method is suitable for all autofocusing approaches that investigate the focus conditions in multiple locations. Our method replaces the insufficient uniform search strategy of the conventional autofocusing algorithms with a suitable optimization tool, i.e., GSS-PI, to find the minimum of the defocus function thus limiting the number of probed locations. The downside of GSS-PI is that to guarantee convergence to the focal location, the defocus function should be unimodal in the declared search range. This limitation can be potentially overcome by performing the search independently in several subintervals or by carefully selecting the region of interest (for the case of nonoverlapping samples). The computational gain of the proposed algorithm offers a promising perspective of realizing demanding autofocusing tasks (large search range, high required accuracy, numerous samples) in a reasonable time.

The proposed GSS-PI approach was applied to three diverse autofocusing cases, providing the computational gain in a range from 136-fold to 7-fold, depending on the complexity of the autofocusing task. Crucially, the achieved speed-up did not deteriorate the autofocusing accuracy. The three examples used for validation of the proposed approach are of special relevance for coherent imaging and metrology. The first one relates to a small uncontrolled defocus appearing for unwanted sources (movement of the sample, system vibrations, etc.) when imaging a biological sample. This example demonstrates the possibility of achieving active focusing at a low consuming time to always set the best in-focus image of the sample for every frame. The second case addresses lensless DHM where imaging the sample to its best in-focus plane by numerical propagation is crucial to visualize and characterize the sample in a vast range of applications such as, for instance, tracking flowing particles, sperm cells sorting, and biological monitoring of living cells. Lastly, the third example deals with fine-tuning to get the best in-focus image for the same sample with different tilted beam illuminations. This is an example of application of the autofocusing for the system calibration in holographic tomography before assembling the final volumetric image.

**Funding.** This work has been partially funded by the National Science Center Poland (SONATA 2020/39/D/ST7/03236), the grant funded by the Scientific Council for the Discipline of Automatic Control, Electronics and Electrical Engineering (Warsaw University of Technology), BIOTECHMED-1 project granted by Warsaw University of Technology under the program Excellence Initiative: Research University (ID-UB) and Foundation for Polish Science FNP (START 2020). Also, part of this work has been supported by the Ministerio de Economía y Competitividad under the project FIS2017-89748-P and the project NanoTech4ALS no. 12/EuroNanoMed/2016, funded under the EU FP7 M-ERA.NET program.

**Acknowledgments.** We thank Barbara Łukomska, Katarzyna Drela and Hanna Trusiak from NeuroRepair Department, Mossakowski Medical Research Institute, Polish Academy of Sciences, Warsaw, Poland for preparing the cells studied in Fig. 4.

**Disclosures.** The authors declare no conflicts of interest.

**Data availability.** Data underlying the results presented in this paper are not publicly available at this time but may be obtained from the authors upon reasonable request.